# IP determination and 1+1 REMPI spectrum of SiO at 210-220 nm with implications for SiO$^+$ ion trap loading

Patrick R. Stollenwerk, Ivan O. Antonov, and Brian C. Odom

Department of Physics and Astronomy, Northwestern University,

2145 Sheridan Road, Evanston IL 60208, USA

**Abstract**

The 1+1 REMPI spectrum of SiO in the 210-220 nm range is recorded. Observed bands are assigned to the A-X vibrational bands (v"=0-3, v'=5-10) and a tentative assignment is given to the 2-photon transition from X to the n=12-13 [$X^2\Sigma^+$,$v^+$=1] Rydberg states at 216-217 nm. We estimate the IP of SiO to be 11.59(1) eV. The SiO$^+$ cation has previously been identified as a molecular candidate amenable to laser control. Our work allows us to identify an efficient method for loading cold SiO$^+$ from an ablated sample of SiO into an ion trap via the (5,0) A-X band at 213.977 nm.

**Introduction**

Interest in spectroscopy of SiO and SiO$^+$ was initially stimulated by astrochemistry. After detection of vibrational bands of SiO in stellar atmospheres[1] a search for electronic transitions followed. A number of studies investigated transitions of SiO in the UV region[2, 3]. The ionization potential (IP) of SiO was measured by means of electron impact[4], spectroscopy of Rydberg states[5], photoelectron spectroscopy (PES)[6], and direct VUV photoionization[7]. SiO+ bands were observed in 1943[8] but at that time misidentified as SiN. Later, this misunderstanding was resolved[9] and several electronic states were mapped by means of PES[6], absorption spectroscopy[9, 10] and laser-induced fluorescence (LIF)[11]. Several *ab initio* calculation studies predicted[12-14] rich excited electronic state structure of SiO$^+$. Ion chemistry of Si$^+$/O$_2$ system and physics of ablation of SiO was studied in several works[15-17], stimulated by interest in Si clusters and vapor deposition.

Our interest in SiO and SiO$^+$ is primarily motivated by its properties that make it amenable to laser control of internal and external degrees of freedom [18, 19]. Control of molecules offers new applications in precision measurements of fundamental constants and ultracold chemistry[20-22]. The quantum state preparation necessary for coherent control is a serious technical challenge for molecules as rotational and vibrational degrees of freedom significantly complicate the internal structure compared to atoms. However, the $B^2\Sigma^+$-$X^2\Sigma^+$ electronic transition of SiO$^+$ has highly diagonal Franck-Condon factors and well-separated P and R branches, which make it suitable for broadband rotational cooling[23]. In order to design a laser cooling scheme for SiO$^+$, accurate knowledge of branching fractions of SiO$^+$ radiative relaxation to low lying vibrational and electronic states is necessary. A previous measurement, done by our group, of emission branching factors to high vibrational bands of $B^2\Sigma^+$-$X^2\Sigma^+$ found branching to the (1,0) band at 3% relative to the (0,0) band, while branching to higher vibrational levels and low-lying $A^2\Pi$ state could not be detected[24]. We determined that further limits were necessary before we could apply a laser cooling scheme.

To enhance the sensitivity of SiO$^+$ spectroscopy measurements, we designed a system that can trap large numbers of SiO$^+$ ions in a linear Paul trap and perform dispersed LIF measurements to extract branching fractions. The work in this paper was motivated by the need for a reliable, rapid, and pure source of SiO$^+$. Efficient and selective loading of SiO$^+$ into a Paul trap can be reliably achieved by means of photoionization. SiO has a relatively high ionization potential (>11 eV), therefore a multiphoton process is needed to photoionize it with commonly available laser light sources. The band system of $A^1\Pi - X^1\Sigma^+$ transitions have previously been studied via LIF[25] and absorption measurements[26]. The energy of these transitions lie approximately halfway between the ground state and the IP of SiO and could thus be used for 2-photon 1-color photoionization; however, no photoionization study of this system exists. This work reports the 1+1 REMPI (resonantly enhanced multiphoton ionization) spectrum of SiO in the 210-220 nm range. The spectrum revealed several new features due to highly excited Rydberg states of SiO and resulted in accurate determination of the IP of SiO.

**Experimental**

All REMPI measurements occurred under UHV (5·$10^{-10}$ Torr) conditions with the aid of a home built linear Paul trap for ion storage and a channel electron multiplier (CEM) for ion detection. A pressed pellet of SiO, situated below the center of the Paul trap, was laser ablated by the second harmonic of a pulsed YAG laser. REMPI was subsequently performed on the ablation plume inside the trapping volume. After accumulating ions from a predetermined number of REMPI and ablation pulses, the trap RF voltage was ramped down, ejecting the trapped ions. The ejected ions are then detected by the CEM.

The experimental setup is shown in Fig 1. Contained within the vacuum chamber is the Paul trap (2Πx3.6 MHz trap frequency, $r_0$=3 mm, $z_0$=7.29 mm, $V_{pp}$=660 V, κ=0.22, $V_{ec}$=~1 V), an ablation target of pressed SiO located ~1 cm below trap center, and the CEM, which is kept at 3 kV throughout the duration of the experiment. Outside the vacuum chamber is a function generator for ramping down the trap RF voltage, the laser probe used for REMPI (Ekspla NT342/1/UVE OPO, 10 Hz rep rate, 4.2 ns pulse width, 4 $cm^{-1}$ linewidth) and a photodiode along the beam path to normalize output power. The ablation source consists of a 532 nm pulsed source (Continuum Minilite II, 10 Hz rep rate, 3-5 ns pulse width) attenuated by a λ/2 waveplate and a polarizing beam splitter to ~0.5 mJ, passed through a motor controlled window rotated to move the beam, and focused onto the sample with a 150 mm lens. As the sample is unable to be moved in the vacuum chamber, the rotating window allows the beam to be walked in a circle and ablate a fresh spot on the surface each pulse. This acts to stabilize the signal which would otherwise continuously decay as the ablation laser would dig a hole over time.

The REMPI laser pulse was delayed by 90 μs with respect to the ablation laser. The REMPI signal was not found to vary significantly over a delay range of ~50 μs. The REMPI beam was lightly focused into the trapping volume with a 200 mm lens and typically averaged a pulse energy of several hundred μJ reaching the ablation plume. The REMPI laser was scanned from 210-220 nm in step sizes of 0.025 nm. At each wavelength, the ablation laser and REMPI laser pulsed 5-20 times at 10 Hz before the RF trapping voltage was ramped down in 0.5 ms and the accumulated ion products detected on the CEM were counted. Care was taken to avoid saturation effects due to filling the ion trap. A boxcar integrator was used to record the average photodiode voltage observed due to the REMPI laser pulses in order to normalize the signal size. This sequence was repeated several times to accumulate statistics before the wavelength was stepped forward. In a given wavelength scanning range, the wavelengths were also stepped in the reverse direction to observe any drifting behaviors found in the signal size. Using a rotating window in the ablation laser path was crucial in eliminating decay in the signal during the reverse of the scan.

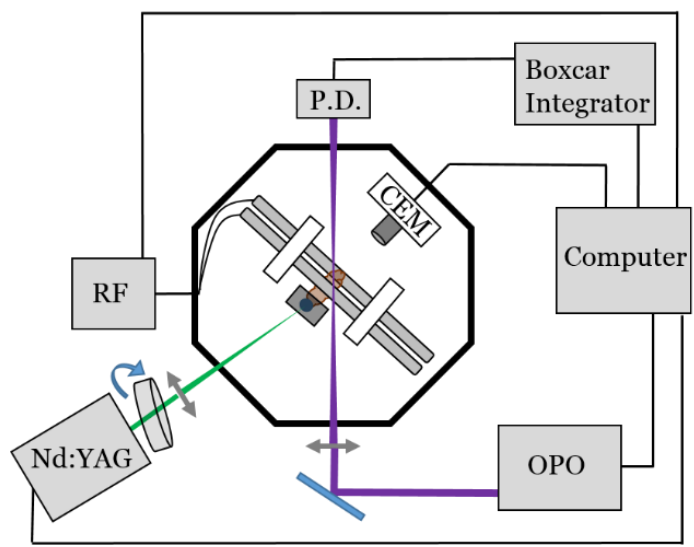

Figure 1. Schematic diagram of the experimental setup.

**Calculations**

The IP of SiO was calculated at CCSD(T)/CBS level. The potential energy curves of the ground electronic states of SiO and SiO$^+$ were calculated at EOM-CCSD level and fitted to a Morse potential to determine $R_e$. The values obtained for the neutral SiO (Re=1.5094 Å) and for SiO$^+$ (Re=1.5166 Å) are in very good agreement with experimentally measured values[27, 28]. CBS extrapolation was achieved by fitting CCSD(T)/ aug-cc-pVnZ (n=2-5) ground electronic state energies of neutral and cation SiO with the mixed exponential/Gaussian formula[29]. The adiabatic IP value was obtained by subtracting the CCSD(T)/CBS energy of SiO from that of SiO$^+$ and correcting for the zero-point energies. The resulting value was IP = 11.6138(1) eV where the number in parenthesis is the 1-σ of the exponential fit parameter.

**Notation**

The following notation is used throughout the paper. Rotational and vibrational levels of the ion are followed by the uppercase + sign, such as $N^+$ or $v^+$. Quantum numbers of the ground state of SiO, $X^1\Sigma^+$ are followed by double prime, e.g. J" and v". Energy levels of $A^1\Pi$ state of the SiO are referred with a single prime, e.g. J' and v'. We chose to use the total rotational quantum number J for the neutral SiO and nuclear rotational quantum number $N^+$ for SiO$^+$ to avoid dealing with half-integer $J^+$ values. Rydberg states of SiO are denoted with (Rydberg electron configuration) state term symbol [threshold ion level] notation. Here a configuration of the Rydberg electron in parenthesis, such as (nsσ), is followed by the state term symbol, such as $^1\Sigma^+$ and the level of the SiO$^+$ ion core, such as [$X^2\Sigma^+$,$v^+$=1], in square brackets.

**Results**

Graph 1 presents the 1+1 REMPI spectrum of neutral SiO at 210-220 nm. The upper black trace is experimental data; colored lines below are simulated transitions. The spectrum has several similarly looking red-degraded bands in the 210-218 nm range. At 216 nm there is a complex band with multiple narrow lines, and after 218 nm the signal is small and mostly unresolved. The rotational lines could not be resolved since the bandwidth of the OPO light source that we used was ~4 cm$^{-1}$, and the rotational constant of SiO is ~0.72 cm$^{-1}$. Therefore the bands observed represent rotational contours of vibronic transitions. A number of bands at 210-218 nm were assigned as (v',v") A-X vibrational bands (v"=0-3, v'=5-10) (see Table 1 and Graph 1); they are red-degraded since the rotational constant of the A state is much smaller than that of the X state (~0.62 cm$^{-1}$ vs ~0.72 cm$^{-1}$ for v=0). The 0-5 rotational band contour at ~214 nm looks different from other X-A bands and the intensity of this band is not reproduced by simulation. The reason for this discrepancy is that the 2-photon excitation on the 0-5 transition brings SiO just above the IP level and direct ionization is not possible for high rotational levels (see discussion for details).

The band at 216-217 nm looks different from the A-X bands and was assigned to another electronic state of SiO. It is likely produced by 2-photon excitation of Rydberg states converging to the v$^+$=1 level of the X$^2\Sigma^+$ electronic state of SiO$^+$. The 2-photon excitation of the (1,0) band of Ry, n=12-13 (X$^2\Sigma^+$,v$^+$=1 ion core) – X $^1\Sigma^+$ is enhanced by absorption of the first photon on the intermediate v'=6 level of the A$^1\Pi$ state of SiO. The 2-photon excitation of the ground state allows transitions to Rydberg states of Σ, Π and Δ symmetry. In the simulation we used the 3 $^1\Sigma^+$ Rydberg states to fit the band shape and the resulting fit has a good quality; however, this assignment is at best tentative and higher precision data is needed to properly assign states involved in the 216-217 nm band.

To summarize, the observed bands were assigned to 1+1 REMPI via the (v',v") A$^1\Pi$ - X$^1\Sigma^+$ intermediate state, where v'=5-9 and v''=0-3, and to the 2-photon excitation of the Rydberg states of SiO with main quantum number n=12-13 converging to X$^2\Sigma^+$,v$^+$=1. The list of observed transitions is shown in Table 1, along with vibrational assignments, FCFs and the energy difference between the energy of the molecule after 2-photon absorption and IE$_{SiO}$. The band intensities of observed X-A vibronic transitions were well reproduced by Franck-Condon factors for X-A and A-X$^+$ systems. Band intensities of the 2-photon transitions to Rydberg states were fitted as independent parameters. The best fit was obtained when both rotational and vibrational temperatures were equal to T=1000 K.

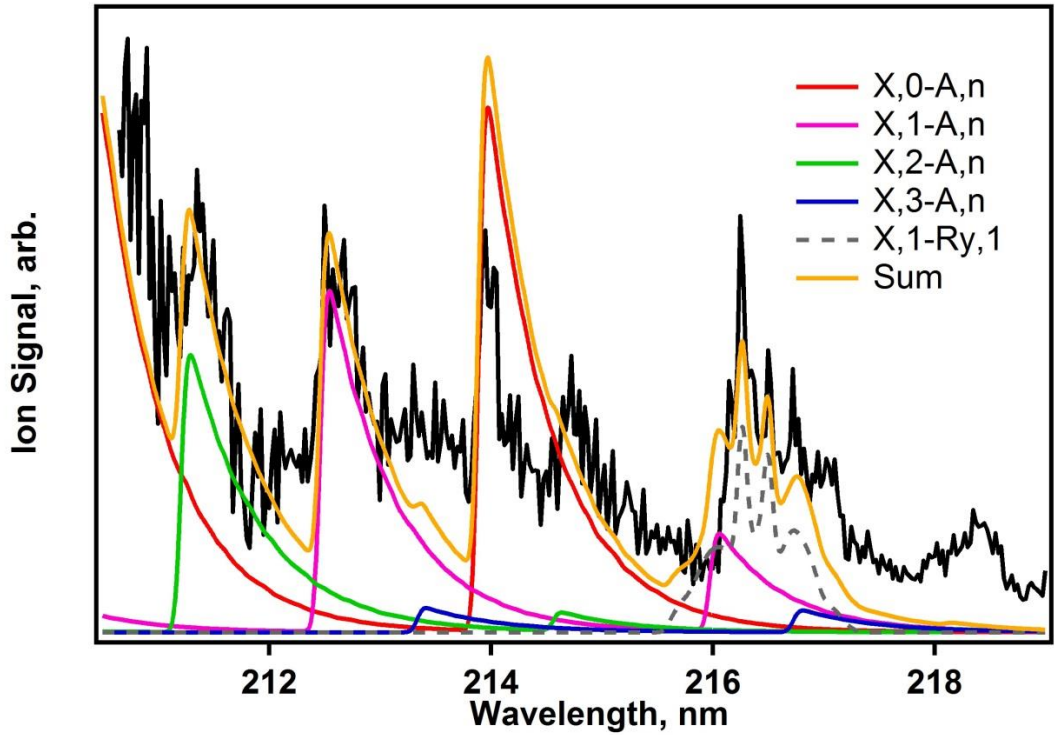

Graph 1. SiO 1+1 REMPI spectrum as a function of UV wavelength at 210-220 nm. Black trace is experimental data and colored traces are simulated transitions (see legend)

| Wavelength (nm) | Assignment | FCF, x10$^{-3}$ | ΔE, cm$^{-1}$ |
|---|---|---|---|
| 210.47* | (6-0), A-X | 3.6 | 1579 |
| 211.29* | (9-2), A-X | 15.9 | 3671 |
| 212.54* | (7-1), A-X | 7.2 | 1882 |
| 213.43* | (10-3), A-X | 16.4 | 3951 |
| 213.98* | (6-0), A-X | 3.6 | 23 |
| 214.66* | (8-2), A-X | 5.7 | 2185 |
| 216.06* | (6-1), A-X | 3.8 | 349 |
| 216.26** | $(s\sigma)^1\Sigma^+[X^2\Sigma^+, v^+=1]$-X,v"=1 | - | ~240 |
| 216.49** | $(p\sigma)^1\Sigma^+[X^2\Sigma^+, v^+=1]$-X,v"=1 | - | ~140 |
| 216.58** | $(d\sigma)^1\Sigma^+[X^2\Sigma^+, v^+=1]$-X,v"=1 | - | ~100 |
| 216.81* | (9-3), A-X | 10.9 | 2489 |
| 218.19* | (7-2), A-X | 1.6 | 676 |

Table 1. Assignment of 1+1 REMPI spectrum of SiO. * - 1+1 REMPI via A-X system, ** - 2-photon excitation of Rydberg states of SiO. FCF are products of Franck-Condon factors for X-A transitions[30, 31] and energy-allowed sum of Franck-Condon factors for ionization from the A state. ΔE is calculated, assuming IE = 11.584 eV.

## Discussion

There are no low-lying singlet electronic states in SiO, and the lowest triplet state is ~33000 cm$^{-1}$ above the ground state. Assuming that SiO molecules are produced by ablation in the ground electronic state (or alternatively, excited molecules are quickly quenched to the ground state by means of collisions or radiation), at 210-220 nm only the transitions with energy hv > (IE$_{SiO}$ – E$_g$ – ΔE$_F$)/2 can result in ionization of SiO. Here IE$_{SiO}$ is the ionization energy, E$_g$ is the ground state (rotational and vibrational) energy and ΔE$_F$ is the depreciation of the SiO IE due to the electric field in the trap. Fig 2a shows energies of the SiO molecule after resonant 2-photon excitation on the A$^1\Pi$ - X$^1\Sigma^+$ transition starting at v"=0-3 level of the ground state of SiO. The energies in this diagram are relative to the minimum of the X$^1\Sigma$ potential energy curve. The dashed grey line is the IP of SiO where we used the IP value of 11.584 eV, measured by Baig and Connerade[5]. To ionize SiO, the two photons should promote it above the IP level. Therefore, SiO molecules starting in v"=0 can be ionized via A-X transition when v'≥5, ionization of v"=1 requires v'≥6. Higher v" levels require correspondingly higher v' levels of the A state of SiO.

The Figure 2 right side plot shows energies of rotational levels of SiO after 2-photon excitation on the (5,0) A – X transition starting at J" = 0-25 rotational levels. The red, green and blue curves with hollow circles denote to excitation via P, Q and R rotational branches of the (5,0) A – X transition. The grey dashed line again is the IP of SiO. It is clear that all initial rotational levels of the 0-5 transition are excited above the IP and in principle can result in ionization. The solid black curve in the blue shaded area shows the energy of the SiO$^+$ ion at X, v$^+$=0 level with N$^+$ = J±1. When the initial rotational state is J" = 7-10 for the P branch, 10-14 for the Q branch and 15-21 for the R branch, the P,Q and R curves lie inside the blue shaded area, which means that direct ionization to N$^+$ = J"±2 rotational level of the ion is possible. Transitions with lower J" values can also result in direct ionization since excess energy is removed by the photoelectron. Transitions with higher J" values cannot produce ions with N$^+$=J±1 directly since there is not enough energy absorbed. Ionization to lower N$^+$ levels is energetically possible but requires removal of one or more quanta of angular momentum by the leaving electron. This could be achieved if a Rydberg state with ion core at N$^+$=J±1 is excited by the second photon and pre-ionizes to form the SiO$^+$ cation with a lower N$^+$ value. This mechanism however will be in competition with radiative and collisional relaxation of the Rydberg state molecules and physical removal of the excited molecules out of the stable trap volume. Therefore, it may result in depletion of intensities of high J" rotational lines in the (5,0) A – X band transition. This is consistent with the spectrum shown in Graph 1 where we observe a narrowed band profile of (5,0) A – X band.

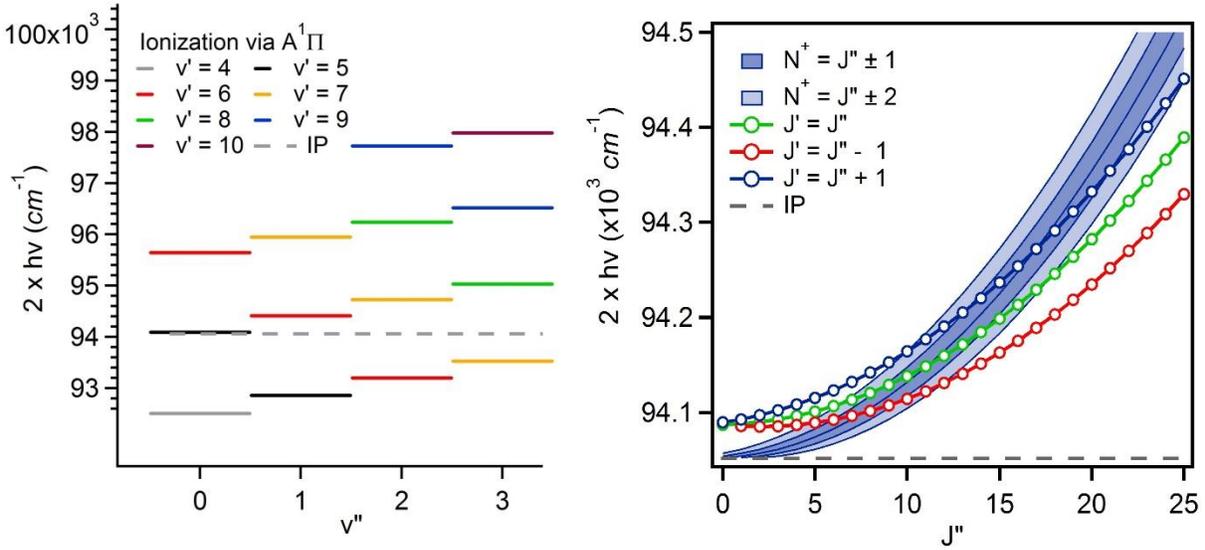

Figure 2. The left plot shows the energy of SiO after 2-photon transition via vibrational bands of A-X transitions. The right plot shows the energies of rotational states after 2-photon excitation of the (5,0) A-X transition.

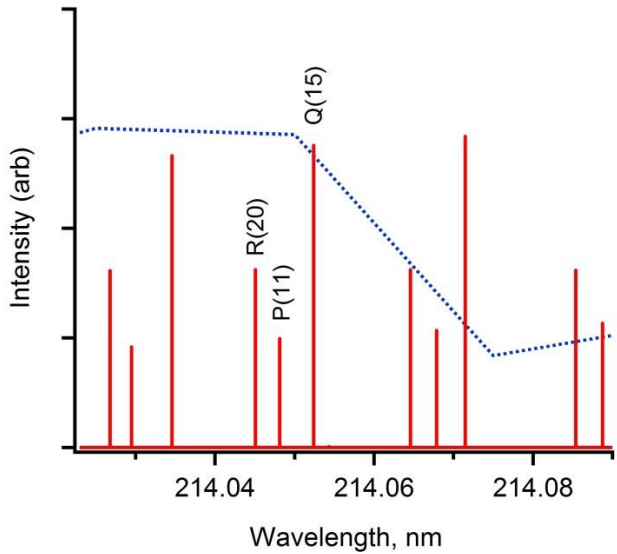

Figure 3 shows a part of the REMPI spectrum with the long wavelength cut-off of the 0-5 X-A band near 214.05 nm overlaid with simulated rotational line positions. The observed beginning of the cut-off corresponds to Q(15), P(11) and R(20) rotational lines which is consistent with prediction made above using Figure 2b for IP = 11.584 eV.

The IP of 11.584 eV estimated from our spectrum has to be modified to account for $\Delta E_F$ -depreciation of ionization energy by electric field in the quadrupole trap. It is known that measured the IP can be reduced by the external electric field[32] by $2\text{-}6*\sqrt{E(V/cm)}$ cm$^{-1}$. Our trap simulations suggest that ions created in an electric field up to 500 V/cm can be trapped in stable trajectories. Therefore, an upper bound for $\Delta E_F$ can be estimated as $6*\sqrt{500(V/cm)}$ = 134 cm$^{-1}$ or ~17 meV, which yields an upper bound for the IP of SiO at 11.601 eV. Assuming that the 17 meV depression of the IP due to the field is the extreme case and given that most molecules that we ionize will be generated under weaker fields, we can estimate that the actual IP is 11.59(1) eV. This value is within error bounds of the most reliable previous measurements[4-7]. Interestingly, the most accurate values obtained with PES[6] (11.61(1) eV) and Rydberg formula fit[5] of the n=4-16 (11.584(11) eV) disagree with each other and our measurement is halfway between those and within error bounds of both of them.

Our theoretical prediction for the IP = 11.6138(1) eV overestimates the experimental value by ~13 meV. It is a good agreement for the CCSD(T)/CBS method which is known to achieve accuracy of 10-20 meV in IP calculations[33]. The discrepancy between measured and calculated values may be due to unaccounted effects, such as relativistic motion or core correlation, and from high level dynamic correlation missing in CCSD(T). Previous calculations by Das et al underestimated the IP by ~0.7 eV, probably because the MRDCI method they used could not treat the dynamic correlation in SiO$^+$ and SiO at the same level.

The band near 216 nm has three sharp features that were identified as Q-branch transitions that correspond to a small change in bond length, i.e. nearly "vertical". While a (6,1) A-X band is predicted to lie near 216 nm, the shape of the 216 nm band observed in this work is very different from the simulated A-X bands. The bond length in SiO, SiO$^+$ ground electronic states and high Rydberg states with [X$^2\Sigma^+$] ion core is very similar. Therefore, it is likely that the 216 nm band is a two-photon excitation to a high lying Rydberg state. The (6,1) A-X band is nearly resonant with absorption of the first photon and therefore acts to increase intensity of the two-photon transition to the Rydberg states. The "vertical" character of the observed bands and the near resonance with (6,1) A-X suggests that the transition originates at v"=1 and excites Rydberg states with an [X$^2\Sigma^+$,v$^+$=1] ion core. Under this assumption, the Rydberg states involved in this transition are located 900-1100 cm$^{-1}$ below the v$^+$=1 level. Baig and Connerade detected[5] the n=12-13 Rydberg states 900-1000 cm$^{-1}$ below the [X$^2\Sigma^+$,v$^+$=0] ionization threshold. A sharp intense Q-branch can be observed for the $\Delta$-$\Sigma$ and $\Sigma$-$\Sigma$ two-photon transitions. Therefore, the 216 nm band is likely due to (ns$\sigma$, np$\sigma$, nd$\sigma$) $^1\Sigma^+$ and (nd$\delta$) $^1\Delta$ Rydberg states with an [X$^2\Sigma^+$,v$^+$=1] ion core and n=12-13.

Loading of SiO$^+$ in a Paul trap with 1+1 REMPI is efficient. Typically several thousand SiO$^+$ ions are loaded per single ablation event followed by REMPI ionization. The most convenient transitions is (5,0) A-X at 213.977 nm – it is intense and SiO$^+$ can be loaded with ~30-300 cm$^{-1}$ internal energy. Therefore, loading of vibrationally or electronically excited SiO$^+$ is avoided and the rotational temperature of loaded ions is significantly lower than that of the ablated SiO. Another advantage of the (5,0) transition is that it is not resonant with the A-X and B-X bands of the NO molecule, lying near 215 nm [34]. NO has a relatively low IP of ~9.26 eV [35] and can be ionized with two photons below 267 nm. Even though ultrahigh vacuum is used in this work and the NOx concentration in the air is <1 ppm, there is enough NO in the vacuum chamber to load a few tens of NO$^+$ cations per laser pulse with by 1+1 REMPI. Avoiding strong NO transitions near 215 nm ensures that NO$^+$ ions are not loaded into the trap.

**Conclusions**

The 1+1 REMPI spectrum of SiO in the 210-220 nm range is recorded. Observed bands were assigned to the A-X band system and 2-photon transitions were tentatively assigned to the n=12-13 [X$^2\Sigma^+$,v$^+$=1] Rydberg states. We have reported loading SiO$^+$ ions in a trap by means of 1+1 REMPI via the X-A transition of neutral SiO. This provides an efficient method for loading an ion trap and allows for starting with vibrationally and electronically cold SiO+, while minimizing rotational energy. We estimate the IP of SiO to be 11.59(1) eV, in agreement with previous measurements.